\documentclass{ws-procs961x669}            % default, citations in superscript
\usepackage{amsfonts,amsmath,amssymb,chapterbib,natbib,graphicx,mathtools,hyperref,bm,calrsfs}

\DeclarePairedDelimiter\floor{\lfloor}{\rfloor}
\hypersetup{
	colorlinks=true,       % false: boxed links; true: colored links
	linkcolor=blue,          % color of internal links
	citecolor=blue,        % color of links to bibliography
}
\usepackage[section]{placeins}

\usepackage{fancybox}
\begin{document}
	\title{Two body dynamics in a quadratic modification of general relativity}
	
	\author{Soham Bhattacharyya}
	
	\address{Max Planck Institute for Gravitational Physics (Albert Einstein Institute)\\ Leibniz Universit\"at Hannover}
	
	\begin{abstract}
		It is shown in this study that deviations from the Einstein-Hilbert action at the quadratic level using a proper analyses and suitable dynamical variables lead to a tiny modification to the post Newtonian equations of motion, and non-GR like behavior at very short length scales. 
	\end{abstract}
	
	%\keywords{Style file; \LaTeX; Proceedings; World Scientific Publishing.}
	
	\bodymatter
	
	\section{Introduction}\label{intro}
	General theory of Relativity (GR) is a field theory, one of the first of its kind, that is classified under classical physics. The validity of the theory ranges from solar system length scales to cosmological length scales. It has passed most of its predictions with flying colors, starting from the extra perihelion precession of Mercury in the inner solar system \cite{Einstein:1915bz}, to gravitational lensing \cite{Will:2014zpa}, to the generation of gravitational waves (GW) \cite{Einstein:1916cc},\cite{Abbott:2016blz,Abbott:2016nmj,Abbott:2017oio,Abbott:2017tlp,Abbott:2017vtc}, and finally to the existence of objects called black holes (BH) \cite{EventHorizonTelescope:2019dse}.\\
	
	One of the first calculations that tested the predictive power of GR was the perihelion precession of Mercury, which is to say that GR gave predictions of all the possible orbital dynamics that was possible to be observed in the early 20th century, and its predictive power remains the same to this date. GR predicted that the \emph{gravitational force} will not always be inverse squared of the distance from the source, and that at shorter length scales, the behavior changes. To be precise, a perturbative treatment of GR, also known as the post Newtonian (PN) formalism\cite{Einstein1949OnTM,Infeld1949,Chandra1965,Chandra1967,Chandra1969I,Chandra1969II,Chandra1970}, reveals that a series of shorter (than inverse squared) ranged velocity dependent forces can have a prominent effect on the dynamics of two objects trapped in their mutual gravitational field (like the Sun and Mercury system). Hence, the universality of Newtonian gravity stops being universal when one goes beyond the \emph{Newtonian bubble}, broadly speaking.\\
	
	However, an uncomfortable question pops up when one considers the PN forces of GR, and keeps on continuing the PN series to shorter and shorter length scales. Or when one considers the collapse of a self-gravitating mass of fluid at the extreme scenario. At every point along the collapse, the gravitational force, including the PN forces, are attractive in nature. Hence, if the fermionic pressure, or the force that makes matter solid, is unable to balance the increasingly powerful (but shorter ranged) forces of gravity, a singularity may occur. This is because in the tug of war of all the forces present in one or multiple gravitating systems, gravity seems to win in some extreme conditions. Such conditions can occur in ultra dense self gravitating systems like the end stages of a star collapsing under its own gravity. If there is no other force that can stop the unyielding pull of gravity, matter eventually collapses fully in on itself. But long before such a full collapse, or singularity, can happen, a special surface forms around the eventual singularity. Such a surface is known as the event horizon, or the surface the forms the termination of all \emph{events} in the Universe that lies outside of the surface. Here \emph{event} refers to the special relativistic notation of space-time events, which are, in the sense of mathematical modelling, points in a four dimensional object known as the manifold, which in turn refers to the complete past, present, and future of the Universe that is being modelled.\\
	
	However, the universality of attraction of gravity may not be the case when one goes beyond the Einstein-Hilbert (EH) action, referred to as GR in literature. In order to demonstrate that, a quadratic deviation to the EH action is considered, which is a subclass of post-Einsteinian theories known as $ f\left(R\right) $ theories of gravity \cite{Starobinsky2007,Nojiri:2006ri}. Such theories are characterized by an extra massive scalar degree of freedom\cite{Capozziello2007,Capozziello2008,Capozziello2010}, in addition to the two massless tensor degrees of freedom.\\
	
	A PN treatment of the densitized field equations of a quadratic extension to GR will be performed in this study that resembles the Landau-Lifshitz formulation of GR. It will be shown that the equations of motion have a rather good limit to GR (almost indistinguishable in fact), and only deviates from the predictions of GR at very high energy scales (where energy scales refer to really large velocities and very high spatial curvatures combined).\\
	
	Geometrized units, $ c\,=\,G\,=\,1 $ will be used in this study unless explicitly mentioned otherwise. The notations of \cite{Futamase2007} will be followed except a few changes in variable and index labeling. The mostly minus sign convention will be followed for the metric, that is $ \left(+,-,-,-\right) $.\\
	
	\section{Layout}
	\subsection{The densitized formulation of GR}
	The gothic metric/Landau-Lifshitz/densitized formulation of GR may be motivated from a particular observation regarding the structure of the second-most condensed form of the Einstein field equations in its covariant form (or the canonical form), that is
	\begin{eqnarray}\label{einsricciform}
	R_{\mu\nu} \,-\, \frac{1}{2}\,g_{\mu\nu}\,R &\propto& T_{\mu\nu}
	\end{eqnarray}
	where the first tensor quantity on the LHS of the above is the Ricci tensor, and the scalar multiplied to the metric tensor in the second term of the LHS is the contraction of the Ricci tensor with the metric, known as the Ricci scalar. The RHS is the classical matter energy-momentum-stress density tensor, which contains certain information (although non-exhaustive) about the content that is separate from the space-time. One notices that the net content of the RHS of (\ref{einsricciform}) is essentially a trace reversed form of the Ricci tensor. Trace reversal implies that the trace of the new tensor that was introduced in the beginning of the 20th century has a trace that is negative of the trace of the Ricci tensor, and is known as the Einstein tensor, which leads to the most condensed form of the field equations of GR, or
	\begin{eqnarray}\label{einseinsform}
	G_{\mu\nu} &\propto& T_{\mu\nu}
	\end{eqnarray}
	where all components of the tensor $ G_{\mu\nu} $ has a one-to-one relationship with all of the corresponding components of the matter energy-momentum-stress density tensor. However, the cost of condensation is the loss of information and usability. As it is, Eq. (\ref{einseinsform}) does not tell one about how planets and stars move and how GWs are radiated. To be able to predict such phenomenon mathematically, one needs to 'dirty their hands'.\\
	
	In classical GR, $ T_{\mu\nu} $ can contain fluids (serving as a model for stars, dust, etc.), or it can contain the energy-momentum-stress density equivalent of electromagnetic fields and charges. The Ricci tensor/scalar are obtained from geometric principles governing the nature of surfaces, volumes, and higher dimensional generalization; which are in turn obtained by contracting the tensor quantity that serves as the model for the space-time 'fabric'\footnote{Although the term fabric is used, it is put under quotation to differentiate from an actual material object, which is something that the physics community thought of as 'Aether' before the famous Michelson-Morley experiment. In conventional notion, material or fabric can be torn or broken; however, a similar 'tearing' of space-time is something that does not seem to happen. Hence one must take the concept of 'fabric' with a pinch of salt.}, or the Riemann tensor $ R_{\mu\sigma\rho\nu} $.  The Riemann tensor in GR is a function of the metric and its derivatives (till second). It has certain symmetries that are given as follows
	\begin{eqnarray}
	R_{\mu\sigma\left(\nu\rho\right)} &=& R_{\left(\mu\sigma\right)\nu\rho}\,=\,0 \label{skew}\\
	R_{\mu\left[\sigma\nu\rho\right]} &=& 0 \label{firstbianchi}\\
	R_{\mu\rho\nu\sigma} &=& R_{\nu\sigma\mu\rho} \label{interchange} \\
	R_{\mu\rho\left[\nu\sigma;\alpha\right]} &=& 0 \label{secondbianchi}
	\end{eqnarray}
	The first, third of the identities, that is Eqs. (\ref{skew}) and (\ref{interchange}), are called skew and interchange symmetries respectively. Whereas Eqs. (\ref{firstbianchi}) and (\ref{secondbianchi}) are known as the first and second Bianchi identities respectively. The second Bianchi identity is particularly useful since it leads to the following identity on the Einstein tensor $ G_{\mu\nu} $
	\begin{eqnarray}
	\nabla^\mu\,G_{\mu\nu} &=& 0, \label{einscons}
	\end{eqnarray}
	where $ \nabla^\mu\,\equiv\,g^{\mu\rho}\,\nabla_\rho $ is the index-raised covariant derivative associated with the metric $ g_{\alpha\beta} $. The above, combined with the proportionality condition (\ref{einseinsform}), lead to the following
	\begin{eqnarray}
	\nabla^\mu\,T_{\mu\nu} &=& 0 \label{EMcons}
	\end{eqnarray}
	which is a covariant conservation law on the classical matter (or electromagnetic fields) energy-momentum-stress density tensor, and using which equation of motion of classical matter itself can be deduced. Hence the famous saying by J.A. Wheeler about Eq. (\ref{einseinsform}): \textit{Spacetime tells matter how move; matter tells spacetime how to curve}, is incomplete without throwing in Eq. (\ref{EMcons}) into the fold.\\
	
	With the above motivations one can define a new tensor given by \cite{LANDAU1975259,LANDAU1975295,LANDAU1975345}
	\begin{eqnarray}
	\mathfrak{g}^{\mu\nu} &=& \sqrt{-g}\,g^{\mu\nu} \label{gothdef}
	\end{eqnarray}
	which utilizes the contravariant metric, or the 'dual' of the covariant metric, whose components actually define the space-time structure. Now the point of taking the dual is to simplify the equations that need to be solved, by replacing the problem at hand with another problem which relates to the original problem via transformation and/or reparametrizations. One also notices the presence of the determinant of the metric tensor through $ \sqrt{-g} $ multiplying $ g^{\mu\nu} $. Under the transformation (\ref{gothdef}), Eq. (\ref{einsricciform}) or (\ref{einseinsform}) can be framed as a slightly less condensed (but more useful) form given as follows
	\begin{eqnarray}
	&&\frac{1}{2\,\left(-g\right)}\,\partial_{\alpha\,\beta}H^{\mu\alpha\nu\beta}\,-\,8\,\pi\,t_{LL}^{\mu\nu} \,=\, 8\,\pi\,T^{\mu\nu}_m \label{llgr} \\
	H^{\mu\alpha\nu\beta} &=& \mathfrak{g}^{\alpha\beta}\mathfrak{g}^{\mu\nu} - \mathfrak{g}^{\alpha\nu}\mathfrak{g}^{\beta\mu} \label{maxRie}\\
	16\,\pi\,(-g)\,t_{\mathrm{LL}}^{\alpha \beta} &=& \partial_{\alpha }\mathfrak{g}^{\mu \nu } \partial_{\beta }\mathfrak{g}^{\alpha \beta } -  \tfrac{1}{4} \mathfrak{g}^{\alpha \beta } \mathfrak{g}_{\gamma \zeta } \mathfrak{g}_{\theta \iota } \mathfrak{g}^{\mu \nu } \partial_{\alpha }\mathfrak{g}^{\gamma \theta } \partial_{\beta }\mathfrak{g}^{\zeta \iota }  \nonumber \\ 
	&& + \tfrac{1}{2} \mathfrak{g}_{\gamma \zeta } \mathfrak{g}_{\theta \iota } \mathfrak{g}^{\mu \alpha } \mathfrak{g}^{\nu \beta } \partial_{\alpha }\mathfrak{g}^{\gamma \theta } \partial_{\beta }\mathfrak{g}^{\zeta \iota } + \tfrac{1}{8} \mathfrak{g}^{\alpha \beta } \mathfrak{g}_{\gamma \zeta } \mathfrak{g}_{\theta \iota } \mathfrak{g}^{\mu \nu } \partial_{\alpha }\mathfrak{g}^{\gamma \zeta } \partial_{\beta }\mathfrak{g}^{\theta \iota } \nonumber\\
	&&-  \tfrac{1}{4} \mathfrak{g}_{\gamma \zeta } \mathfrak{g}_{\theta \iota } \mathfrak{g}^{\mu \alpha } \mathfrak{g}^{\nu \beta } \partial_{\alpha }\mathfrak{g}^{\gamma \zeta } \partial_{\beta }\mathfrak{g}^{\theta \iota } -  \partial_{\alpha }\mathfrak{g}^{\mu \alpha } \partial_{\beta }\mathfrak{g}^{\nu \beta } + \mathfrak{g}^{\alpha \beta } \mathfrak{g}_{\gamma \zeta } \partial_{\alpha }\mathfrak{g}^{\mu \gamma } \partial_{\beta }\mathfrak{g}^{\nu \zeta } \nonumber \\ 
	&& + \tfrac{1}{2} \mathfrak{g}_{\alpha \beta } \mathfrak{g}^{\mu \nu } \partial_{\gamma }\mathfrak{g}^{\beta \zeta } \partial_{\zeta }\mathfrak{g}^{\alpha \gamma }  -  \mathfrak{g}_{\beta \gamma } \mathfrak{g}^{\nu \alpha } \partial_{\alpha }\mathfrak{g}^{\gamma \zeta } \partial_{\zeta }\mathfrak{g}^{\mu \beta }  -  \mathfrak{g}_{\beta \gamma } \mathfrak{g}^{\mu \alpha } \partial_{\alpha }\mathfrak{g}^{\gamma \zeta } \partial_{\zeta }\mathfrak{g}^{\nu \beta } \nonumber\\
	\end{eqnarray}
	The definition of the Landau-Lifshitz pseudo-tensor $ t^{\mu\nu}_{LL} $ deviates slightly from the usual definition found in literature since all the metrics that are used to raise or lower the indices have been transformed into the gothic metric. The covariant gothic metric is defined as
	\begin{eqnarray}
	\mathfrak{g}_{\mu\nu} &=& \frac{g_{\mu\nu}}{\sqrt{-g}} \label{covdgothmetric}
	\end{eqnarray}
	One notices that the rank 4 tensor $ H^{\mu\alpha\nu\beta} $ has the same symmetries as the Riemann tensor, and is proportional to the Riemann tensor corresponding to the maximally symmetric Riemann tensor. The maximally symmetric Riemann tensor is a solution of the field equations of GR when the space-time is homogeneous and isotropic everywhere. Deviation from such a space-time with the maximum number of symmetries is marked by the appearance of a coordinate or a gauge dependent tensor (or a pseudo-tensor) $ t^{\mu\nu}_{LL} $, also known as the Landau-Lifshitz energy-momentum-stress density pseudo-tensor. $ t^{\mu\nu}_{LL} $ is a quantification of the energy, momentum, and stress that is inherent in the space-time, and leads to a solutions that are less symmetric than the completely homogeneous and isotropic space-time one started off with. Such solutions are perturbative in nature and formed the early basis for the Post-Newtonian formalism in GR \cite{Einstein1949OnTM,Infeld1949,Chandra1965,Chandra1967,Chandra1969I,Chandra1969II,Chandra1970}. The perturbation is defined about the Minkowski space-time as
	\begin{eqnarray}
	\mathfrak{h}^{\mu\nu} &=& \eta^{\mu\nu}\,-\,\mathfrak{g}^{\mu\nu}. \label{gothtensorpert}
	\end{eqnarray}
	The above tensor is projected on a Cartesian coordinate frame of a freely falling observer by choosing the following gauge condition
	\begin{eqnarray}
	\partial_\mu\,\mathfrak{h}^{\mu\nu} &=& 0 \label{harmonicgauge}
	\end{eqnarray}
	which not only simplifies the field equations but also puts it in a form that offers a more intimate insight into the nature of space-time and its relationship to classical matter. They are given by the following
	\begin{eqnarray}
	\Box \mathfrak{h}^{\mu\nu} &=& -16\,\pi\,\Lambda^{\mu\nu} \label{LLh} \\
	\Lambda^{\mu\nu}&=& \Theta^{\mu\nu} + \partial_{\alpha\,\beta}\chi^{\mu\nu\alpha\beta} \label{effemp} \\
	\Theta^{\mu\nu} &=& \left(-g\right) \left(T^{\mu\nu} \,+\, t_{LL}^{\mu\nu} \,-\, t_H^{\mu\nu}\right)\label{effem}\\
	\chi^{\mu\nu\alpha\beta} &=& \frac{1}{16\,\pi} \left(\mathfrak{h}^{\alpha\nu}\,\mathfrak{h}^{\beta\mu} \,-\, \mathfrak{h}^{\alpha\beta}\,\mathfrak{h}^{\mu\nu}\right) \label{surfterm}
	\end{eqnarray}
	where a proportionality constant of $ 16\,\pi $ has now been used, which is the case when one uses geometrized units ($ G\,=\,c\,=\,1 $). The rank 4 tensor $ \chi^{\mu\alpha\nu\beta} $ with Riemann symmetry, being a total derivative term, can be turned into a boundary term using Gauss's law, and can be discarded as the source of the wave equation (\ref{LLh}) when a particular choice of the Green's function is made for the flat-space Laplacian/wave operator $ \Box $. One immediately notices that the dynamical variable now follows the well studied massless wave equation, which has standard solutions. It is also to be noted that if one chooses the effective energy-momentum-stress density tensor to be the sum of both the matter and the pseudo part, the dynamical variable has a one-to-one relationship with the matter density. The Landau-Lifshitz or the gothic metric formalism is hence often called as the densitized formalism, as it allows the introduction of a variable that can be taken as the space-time density itself, in contrast to the regular metric tensor which has a very complicated relationship with the energy-momentum-stress density tensor. The covariant conservation identity in Eq. (\ref{EMcons}) in this 'reduced' formalism, when added to the Landau-Lifshitz pseudo-tensor, along with the harmonic gauge condition (\ref{harmonicgauge}), reduces to the following
	\begin{eqnarray}
	\partial_\nu\,\Lambda^{\mu\nu} &=& 0. \label{EMconspartial}
	\end{eqnarray}
	
	Now to come back to the motivation for mentioning trace-reversal at the beginning of the section; one may choose to study perturbative GR by simply linearizing the metric tensor about the Minkowski space-time itself, that is
	\begin{eqnarray}
	g^{\mu\nu} &=& \eta^{\mu\nu}\,-\,h^{\mu\nu} \label{metricpert}
	\end{eqnarray}
	At the linear order the gothic metric perturbation and the metric perturbation are related as follows (using Jacobi's identity)
	\begin{eqnarray}
	\mathfrak{h}^{\mu\nu} &=& h^{\mu\nu}\,-\,\frac{1}{2}\,\eta^{\mu\nu}\,h \label{gothhtometrich}\\
	\mathfrak{h} &=& -h \label{tracerelation}
	\end{eqnarray}
	where $ h $ and $ \mathfrak{h} $ are the trace of the metric perturbation and the gothic metric perturbation respectively. Eq. (\ref{gothhtometrich}) is used as a standard textbook solution for modeling the propagation of gravitational waves in empty space-times. Multiplying $ \sqrt{-g} $ to the dual of the metric tensor is also similar to the multiplication of the same factor to the Ricci scalar (other than making the volume element covariant) to make the Einstein-Hilbert Lagrangian density, which has the same dimensions as any other matter Lagrangian density which may appear at the action level along with the Einstein-Hilbert Lagrangian density, that is
	\begin{eqnarray}
	S &=& \int_{\mathcal{M}}\,\sqrt{-g}\,d^4x\,\left(\frac{R}{16\,\pi}\,+\,\mathcal{L}_{matter}\,+\,\mathcal{L}_{electromagnetism}\,+\,...\right) \label{GRaction}
	\end{eqnarray}
	\subsection{The densitized formulation of $ f\left(R\right) $ theories of gravity}
	Like any other field theory, the class of theories known as $ f\left(R\right) $ theories of gravity can be derived from the following action
	\begin{eqnarray}
	S &=& \int_{\mathcal{M}}\,\sqrt{-g}\,d^4x\,\left(\frac{f\left(R\right)}{16\,\pi}\,+\,\mathcal{L}_{matter}\,+\,\mathcal{L}_{electromagnetism}\,+\,...\right) \label{fRaction}
	\end{eqnarray}
	However, in the current study for simplicity, we will not be including electromagnetic fields or any other 'exotic' fields. The field equations can be derived by varying the action with respect to the metric, and its higher derivatives, and by enforcing all variations vanish at the boundaries of the integration, which is well suited to study the classical problem at hand. The field equations so obtained are as follows\cite{Capozziello2010}
	\begin{eqnarray}
	&&f'\left(R\right)\,R_{\mu\nu}\,-\,\frac{1}{2}\,f\left(R\right)\,g_{\mu\nu}\,-\,\left[\nabla_\mu\,\nabla_\nu\,-\,g_{\mu\nu}\,\Box\right]\,f'\left(R\right) \,=\, 8\,\pi\,T_{\mu\nu} \label{fReom}\\
	&&f'\left(R\right) \,=\, \frac{df\left(R\right)}{dR} \label{fprimedef}
	\end{eqnarray}
	where $ \nabla_\mu $, as usual, is the covariant derivative associated with the metric $ g_{\mu\nu} $. The above field equations can be written in two different forms (if one intends to use the machinery of GR): one modifies the Newtonian constant (which has been put to unity in the current study), and the other preserves it. The former is written as\cite{Capozziello2010}
	\begin{eqnarray}
	G_{\mu\nu} &=& \frac{8\,\pi}{f'}\,\left(T_{\mu\nu}\,+\,T^{eff}_{\mu\nu}\right) \label{fReomGmod} \\
	T^{eff}_{\mu\nu} &\equiv& \frac{1}{2}\,\left(f\,-\,R\,f'\right)\,g_{\mu\nu}\,+\,\left(\nabla_\mu\,\nabla_\nu\,-\,g_{\mu\nu}\,\Box\right)f' \label{TeffGmod}
	\end{eqnarray}
	where $ \Box\,\equiv\,g^{\alpha\beta}\,\nabla_\alpha\,\nabla_\beta $. The latter version has the following form
	\begin{eqnarray}
	G_{\mu\nu} &=& 8\,\pi\,T_{\mu\nu}\,+\,\tilde{T}^{eff}_{\mu\nu} \label{fReomGnotmod} \\
	\tilde{T}^{eff}_{\mu\nu} &=& \left(1\,-\,f'\right)\,R_{\mu\nu}\,+\,\frac{1}{2}\,\left(R\,-\,f\right)\,g_{\mu\nu}\,+\,\left(\nabla_\mu\,\nabla_\nu\,-\,g_{\mu\nu}\,\Box\right)\,f' \label{TeffGnotmod}
	\end{eqnarray}
	In this study, only the former, that is Eqs. (\ref{fReomGmod})-(\ref{TeffGmod}) will be considered because of convenience, and the extra terms are less in number. There is also another differential equation that is satisfied by the scalar function $ f' $, which is independent of whatever form is chosen for the effective energy-momentum-stress density tensor, and is obtained by taking the trace of either Eqs. (\ref{fReomGmod}) or (\ref{fReomGnotmod}). It's given by
	\begin{eqnarray}
	3\,g^{\mu\nu}\,\nabla_{\mu}\,\nabla_{\nu}\,f'\,+\,f'\,R\,-\,2\,f = 8\,\pi\,T \label{scalfield}
	\end{eqnarray}
	and can be regarded as the dynamics of some effective scalar field. The transition of $ f' $ from an 'effective' to an actual scalar degree of freedom will be made clear in a few paragraphs.\\
	
	Before trying to obtain an equivalent of Eq. (\ref{llgr}) one must be made aware of a particular conformal property of $ f\left(R\right) $ theories of gravity, that is under a general conformal transformation of the metric tensor
	\begin{eqnarray}
	\tilde{g}_{\mu\nu} &=& \Omega^2\,g_{\mu\nu}
	\end{eqnarray}
	$ f\left(R\right) $ theories can be written as an \textit{Einstein + massive scalar} system. The $ f\left(R\right) $ action without matter, for example transforms as \cite{Nojiri:2010wj,Magnano:1993bd,Chakraborty:2018ost,Bahamonde:2017kbs,Nashed:2019yto}
	\begin{eqnarray}
	S &=& \int_{\mathcal{M}}\,d^4x\,\sqrt{-g}\,\left(\frac{\tilde{R}}{16\,\pi}\,+\,\frac{1}{2}\,\tilde{g}^{\mu\nu}\,\phi_{,\mu}\,\phi_{,\nu}\,-\,V\left(\phi\right)\right) \label{confaction} \\
	\phi &\equiv& \sqrt{\frac{3}{16\,\pi}}\,\ln\Omega \label{phidef} \\
	V\left(\phi\right) &\equiv& \frac{1}{16\,\pi}\,\frac{f\,-\,R\,f'}{\left(f'\right)^2} \\
	\Omega &\equiv& f'
	\end{eqnarray}
	Where $ \tilde{R} $ is the Ricci scalar of the conformally transformed metric $ \tilde{g}_{\mu\nu} $. This confirms the presence of an extra massive scalar degree of freedom in $ f\left(R\right) $ theories of gravity. In the literature of $ f\left(R\right) $ and scalar-tensor theories of gravity, the `vanilla' metric $ g_{\mu\nu} $ is known as the Jordan frame, whereas the conformally transformed metric $ \tilde{g}_{\mu\nu} $ is called the Einstein frame. From this motivation, one can define a new gothic metric (in the Jordan frame) $ \tilde{g}^{\mu\nu} $ given by
	\begin{eqnarray}
	\tilde{\mathfrak{g}}^{\mu\nu} &=& f'\left(R\right)\,\sqrt{-g}\,g^{\mu\nu} \label{newgothdef}
	\end{eqnarray}
	which frames the field equations (\ref{fReomGmod}) in the following form
	\begin{eqnarray}
	\partial_{\alpha\,\beta} \tilde{H}^{\alpha\mu\beta\nu}  &=& -16\,\pi\,\left(-g\right)\,f'\left(R\right) \,\left(T^{\mu\nu} \,+\, t_{eff}^{\mu\nu}\,+\, \tilde{t}^{\mu\nu}_{LL}\right), \label{fRLL} \\
	16\,\pi\,\left(-g\right)\,f'\,t^{\mu\nu}_{eff} &=& \sqrt{-g}\,\left(f\,-\,R\,f'\right)\,\tilde{\mathfrak{g}}^{\mu\nu} \,+\, \frac{3}{\left(f'\right)^2}\,\left(\tilde{\mathfrak{g}}^{\mu\alpha}\,\tilde{\mathfrak{g}}^{\nu\beta}\,-\,\frac{1}{2}\,\tilde{\mathfrak{g}}^{\mu\nu}\,\tilde{\mathfrak{g}}^{\alpha\beta}\right)\nonumber\\
	&&\times\quad\,\partial_{\alpha}f'\,\partial_{\beta}f' \label{teffgenfR} \nonumber\\
	\end{eqnarray}
	where $ \tilde{H}^{\alpha\mu\beta\nu} $ and $ \tilde{t}_{LL}^{\mu\nu} $ has the same form as that of $ H^{\alpha\mu\beta\nu} $ and $ t^{\mu\nu}_{LL} $, respectively, but with $ \mathfrak{g} $ replaced by $ \tilde{\mathfrak{g}} $. The newly appeared tensor $ t^{\mu\nu}_{eff} $ contains products of only quadratic forms of first derivatives of $ f' $, as seen in Eq. (\ref{teffgenfR}). Owing to the conformal transformation, all the double derivatives of $ f' $ in the tensor field equations cancel away, making $ t^{\mu\nu}_{eff} $ a perfectly usable source for the perturbative post Newtonian expansion. Curiously, the definition Eq. (\ref{newgothdef}) appears as an identity of Palatini formalism of $ f\left(R\right) $ theories of gravity as well, for example in Eq. (17) of \cite{Sotiriou:2008rp}.\\
	
	Now before a post Newtonian sequence of solutions can be found, one needs to choose a particular form of the function $ f\left(R\right) $, which consequently fixes the form of the function $ f'\left(R\right) $ as well, and can be taken for simplicity as
	\begin{eqnarray}
	f\left(R\right) &=& R\,+\,\frac{f''\left(0\right)}{2}\,R^2 \label{fRform}\\
	f'\left(R\right) &=& 1\,+\,f''\left(0\right)\,R \label{fpRform}
	\end{eqnarray}\\
	where the coefficients of expansion $ f'\left(0\right) $ is taken to be unity to recover GR at the $ R\,=\,0 $ limit, and $ f''\left(0\right) $ will be taken to be negative for the course of this article, following \cite{Berry2011,Schmidt1986,Teyssandier1990,Olmo2005}. One defines a tensor field $ \tilde{\mathfrak{h}}^{\mu\nu} $, similar to Eq. (\ref{gothtensorpert}), that propagates on the Minkowski background. One can define that as follows
	\begin{eqnarray}
	\tilde{\mathfrak{h}}^{\mu\nu} &=& \eta^{\mu\nu}\,-\,\tilde{\mathfrak{g}}^{\mu\nu}. \label{newgothtensorpert}
	\end{eqnarray}
	In order to obtain wave like equations of motion where the coordinate variable functions comprising the coordinate 4-vectors follow the conditions
	\begin{eqnarray}
	\Box\,x^{\mu} &=& 0,
	\end{eqnarray}
	one must choose the following gauge/coordinate condition similar to Eq. (\ref{harmonicgauge})
	\begin{eqnarray}
	\partial_\mu\,\tilde{\mathfrak{h}}^{\mu\nu} &=& 0 \label{conformalharmonicgauge}
	\end{eqnarray}
	which will be called as the conformal harmonic gauge. Now similar to the conformally transformed new metric density $ \tilde{\mathfrak{g}}^{\mu\nu} $, one must also define a scalar variable corresponding to a scalar density, say $ \mathfrak{R} $, that follows a massive wave equation (to model the dynamics of the massive scalar degree of freedom), and maintains the $ \left(-g\right) $ relationship with the trace of the net energy-momentum-stress density tensor $ \tilde{\Lambda}^{\mu\nu} $ as well, as in Eq. (\ref{LLh}). One may define that as follows
	\begin{eqnarray}
	\mathfrak{R} &=& \sqrt{-g}\,R. \label{gothriccidef}
	\end{eqnarray}
	Substituting Eq. (\ref{newgothtensorpert}), (\ref{gothriccidef}), and (\ref{conformalharmonicgauge}) in the reduced field equations (\ref{fRLL}), one obtains the following
	\begin{eqnarray}
	\Box \mathfrak{\tilde{h}}^{\mu\nu} &=& -16\,\pi\,\tilde{\Lambda}^{\mu\nu} \label{LLfR}\\
	\tilde{\Lambda}^{\mu\nu} &=& \left(-g\right)\,\left[T^{\mu\nu}_m\,\left(1\,-\,\frac{f''\,\,\mathfrak{R}}{\sqrt{-g}}\right)\,+\,t_{LL}^{\mu\nu}\,+\,t_H^{\mu\nu}\,+\,t_{eff}^{\mu\nu}\right]. \label{effEMSfR}
	\end{eqnarray}
	$ t_{eff}^{\mu\nu} $ is comprised of various products of $ \tilde{\mathfrak{h}}^{\mu\nu} $, $ \mathfrak{R} $, and first derivatives of $ \mathfrak{\tilde{h}}^{\mu\nu} $ and $ \mathfrak{R} $; whose truncated form till the quadratic order of $ f''\left(0\right) $ is given as follows
	\begin{eqnarray}
	&&16\,\pi\,\left(-g\right)\,t^{\mu\nu}_{eff} = \,\frac{f''\,\,\mathfrak{R}^2}{\sqrt{-g}}\,\,\eta^{\mu\nu}\,+\,\frac{3\,\left(f''\right)^2}{4\,\left(-g\right)}\,\left[2\,\left(\eta^{\mu\alpha}\,\eta^{\nu\beta}\,+\,\eta^{\mu\beta}\,\eta^{\nu\alpha}\right.\right.\nonumber\\
	&&\left.\left.\,-\,\eta^{\mu\nu}\,\eta^{\alpha\beta}\right)\,\mathfrak{R}\,\partial_{\alpha}\mathfrak{R}\,\partial_\beta\tilde{\mathfrak{h}}\,+\,\left(\eta^{\mu\alpha}\,\eta^{\nu\beta}\,-\,\frac{1}{2}\,\eta^{\mu\nu}\,\eta^{\alpha\beta}\right)\,\left(4\,\partial_{\alpha}\mathfrak{R}\,\partial_\beta\mathfrak{R}\,+\,\mathfrak{R}^2\,\partial_{\alpha}\tilde{\mathfrak{h}}\,\partial_{\beta}\tilde{\mathfrak{h}}\right)\right]\nonumber\\
	&&\,+\,\mathcal{O}\left[\left(\frac{f''}{\sqrt{-g}}\right)^3\right]\label{teffrel}\\
	&&\tilde{\mathfrak{h}}\,\equiv\,\eta_{\mu\nu}\,\tilde{\mathfrak{h}}^{\mu\nu}
	\end{eqnarray}
	with the following conservation law being satisfied by $ \tilde{\Lambda}^{\mu\nu} $
	\begin{eqnarray}
	\partial_\mu\,\tilde{\Lambda}^{\mu\nu} &=& 0\,\,\,. \label{effemcons}
	\end{eqnarray}
	whereas the new scalar dynamical variable $ \mathfrak{R} $ satisfies the following, as obtained by substituting Eq. (\ref{newgothtensorpert}), (\ref{gothriccidef}), and (\ref{conformalharmonicgauge}) in Eq. (\ref{scalfield})
	\begin{eqnarray}
	&&\Box\,\mathfrak{R}\,+\,\sqrt{-g}\,\gamma^2\,\mathfrak{R} = -8\,\pi\,\left(-g\right)\,\gamma^2\,\tilde{\Lambda} \label{scaldofdyn}\\
	&&\gamma^2 \equiv -\frac{1}{3\,f''\left(0\right)} \label{gammadef}\\
	&&\tilde{\Lambda} = \,T_m\,+\,f''\left(0\right)\,\left\{\frac{T_m\,\mathfrak{R}}{3\,\sqrt{-g}}+\,\frac{1}{8\,\pi\,\left(-g\right)}\left(\frac{\mathfrak{R}^2}{3}\,-\,\partial_\mu\mathfrak{R}\,\partial^\mu\tilde{\mathfrak{h}}\,-\,\frac{1}{2}\,\mathfrak{R}\,\partial_{\mu}\tilde{\mathfrak{h}}^{\alpha\beta}\,\partial^{\mu}\tilde{\mathfrak{h}}_{\alpha\beta}\right.\right.\nonumber\\
	&&\left.\left.\,-\,\frac{1}{4}\,\mathfrak{R}\,\partial_{\mu}\tilde{\mathfrak{h}}\,\partial^\mu\tilde{\mathfrak{h}}\,-\,\frac{1}{2}\,\mathfrak{R}\,\Box\tilde{\mathfrak{h}}\right)\right\}\,+\,\mathcal{O}\left(\left[f''\left(0\right)\right]^2\right)\label{effemtilde}\\
	\\
	&&\,T_m \,=\, \eta_{\mu\nu}\,T_m^{\mu\nu} \label{emscaldef}
	\end{eqnarray}
	
	Before moving on to finding the equations of motion of compact objects, and to conclude this section one may take the LHS of Eq. (\ref{fReomGmod}) to be some modified Einstein tensor $ \mathcal{G}_{\mu\nu} $, that is
	\begin{eqnarray}
	\mathcal{G}_{\mu\nu} &\equiv& f'\left(R\right)\,R_{\mu\nu}\,-\,\frac{1}{2}\,f\left(R\right)\,g_{\mu\nu}\,-\,\left[\nabla_\mu\,\nabla_\nu\,-\,g_{\mu\nu}\,\Box\right]\,f'\left(R\right), \label{modeins}
	\end{eqnarray}
	it can be shown that\cite{Berry2011} the modified tensor $ \mathcal{G}_{\mu\nu} $ satisfies the same covariant conservation condition as the Einstein tensor $ G_{\mu\nu} $, that is
	\begin{eqnarray}
	\nabla^\mu\,\mathcal{G}_{\mu\nu} &=& 0
	\end{eqnarray}
	which further implies
	\begin{eqnarray}
	\nabla^\mu\,T_{\mu\nu} &=& 0.
	\end{eqnarray}
	The above leads to the conclusion that the covariant conservation law holds for $ f\left(R\right) $ theories of gravity in general, and may help in finding a more generalized second Bianchi identity. However, from what is known in the literature (for example in \cite{Berry2011,Koivisto:2005yk}) about the consequence of the covariant conservation law is a peculiar identity on the function $ f'\left(R\right) $, given by
	\begin{eqnarray}
	\left(\Box\,\nabla_{\nu}\,-\,\nabla_{\nu}\,\Box\right)\,f' &=& R_{\mu\nu}\,\nabla^{\mu}\,f'
	\end{eqnarray}
	which is similar to the appearance of the Riemann curvature tensor due to covariant derivative of a vector field not commuting, implying curvature or non-flatness of the space-time. However, the above commuting relation is on a scalar field, implying that some unconventional geometry might be at large here.
	\section{Equations of motion from the covariant conservation law}
	\cite{Futamase1987,Futamase2007} showed that it is possible to obtain the post Newtonian sequence of solutions for beyond the weak field scenario, that is when the post Newtonian parameter squared $ \epsilon^2 $ is proportional to the compactness (or equivalently, the potential at the surface of an ultra-compact object like Neutron stars and small black holes being proportional to the kinetic energy of either bodies). To be precise, in non-geometrized units when
	\begin{eqnarray}
	\epsilon^2 &\sim& \frac{G\,M}{c^2\,a}	
	\end{eqnarray}
	where $ M $ and $ a $ are the total mass and radius of the compact object. In order that strong surface gravity (post Minkowskian) can be incorporated within the post Newtonian (relativistic) framework, a series of coordinate transformations can be made that requires the integration of wave equations like (\ref{LLh}) or (\ref{LLfR}) and (\ref{scaldofdyn}) to be performed in the so-called \textit{body-zone coordinate system}, while not worrying about the effects of strong internal gravity.\\
	
	In order to maintain the compactness of the body to unity at all times, some temporal+spatial scalings are made, with the temporal scaling defining a proper time $ s $ ($ \tau $ in \cite{Futamase1987}) given by
	\begin{eqnarray}
	s &=& \epsilon\,t \label{newtime}
	\end{eqnarray}
	where $ t $ is the coordinate time of the inertial observer. Under the above transformation, the Minkowski metric and the corresponding metric determinant scale as follows
	\begin{eqnarray}
	\eta_{\mu\nu} &=& diag\left(\epsilon^{-2},\,-1,\,-1,\,-1\right) \label{scaledminkowski}\\
	-\det\left(\eta_{\mu\nu}\right) &=& \epsilon^{-2} \label{scaleddetmetric}
	\end{eqnarray}
	which is singular for $ \epsilon\,\rightarrow\,0 $.	The spatial scalings that are going to be used for the integration in the body zone of Eq. (\ref{LLfR}) are given by
	\begin{eqnarray}
	X_L^{i} &\equiv& \frac{x^{i}-z_L^{i}\left(s\right)}{\epsilon^2}. \label{scalcoord}
	\end{eqnarray}
	The scalings in Eq. (\ref{newtime}) and (\ref{scalcoord}) lead to any rank two tensor, or in our case, the energy-momentum-stress density tensor, transforming as
	\begin{eqnarray}
	T_L^{s\,s} &=& \epsilon^2\,T^{t\,t} \,\sim\, \mathcal{O}\left(\epsilon^{-2}\right) \label{emscalss} \\
	T_L^{s\,i'} &=& \epsilon^{-1}\,T_L^{t\,i} \,\sim\, \mathcal{O}\left(\epsilon^{-5}\right) \label{emscalsi}\\
	T_L^{i'\,j'} &=& \epsilon^{-4}\,T^{i\,j}_L \,\sim\,\mathcal{O}\left(\epsilon^{-8}\right), \label{emscalij}
	\end{eqnarray}
	with the un-primed coordinates representing the unscaled coordinates. Now the above scalings, i.e. Eqs. (\ref{newtime})-(\ref{scalcoord}), are made in a coordinate system that is stationary with respect to the origin or the center of mass of the $ L-th $ compact body, denoted by an external observer with the position 3-vector $ z_L^i\left(s\right) $, and the 3-velocity $ v^i_L $ defined as
	\begin{eqnarray}
	v^i_L &=& \frac{d\,z_L^i\left(s\right)}{ds} \label{comveldef}
	\end{eqnarray}
	Hence the next (and final) transformation requires the definition of a co-moving frame (but not co-rotating, since boost and rotation in the wrong order may lead to unnecessary Thomas precession), moving along an ultra compact object whose spin can be ignored. The energy-momentum-stress density tensor of just the matter content of the compact object transforms from an inertial frame to the near-zone or the co-moving frame as follows
	\begin{eqnarray}
	T_L^{s\,s} &=& T_L^{s\,s} \label{emfmss}\\
	T_L^{s\,i} &=& \epsilon^2 T_L^{s\,i'} + v_L^{i} T_L^{s\,s} \label{emfmsi} \\
	T_L^{i\,j} &=& \epsilon^4 T_L^{i'\,j'} + 2\epsilon^2 v_L^{\left(i\right.} T_L^{\left.j'\right) s} + v_L^i \,v_L^j T_L^{s\,s} \label{emfmij}
	\end{eqnarray}
	Then at the leading order the post Newtonian solutions for both GR and $ f\left(R\right) $ theories of gravity, as seen by the inertial observer, are given by
	\begin{eqnarray}
	\mathfrak{h}_\mathcal{B}^{s\,s} &=&  4\, \epsilon^4\, \sum_{L=1,\,2} \, \left(\frac{P^s_L}{r_L} + \epsilon^2\,\frac{D^k_L\,r^k_L}{r_L^3} \right) + \mathcal{O}\left(\epsilon^{8}\right) \label{fhssmul} \\
	\mathfrak{h}_\mathcal{B}^{s\,i} &=& 4 \, \epsilon^4 \sum_{L=1,\,2} \, \left(\frac{P_L^i}{r_L} + \epsilon^2\,\frac{J_L^{k\,i} r_L^k}{r_L^3}\right) + \mathcal{O}\left(\epsilon^8\right) \label{fhsimul} \\
	\mathfrak{h}_\mathcal{B}^{i\,j} &=& 4 \, \epsilon^2 \, \sum_{L=1,\,2} \left(\frac{Z_L^{i\,j}}{r_L} + \epsilon^2 \, \frac{Z_L^{k\,i\,j} \, r_L^k}{r_L^3}\right) + \mathcal{O}\left(\epsilon^6\right) \label{fhijmul}
	\end{eqnarray}
	In the current formalism, the above components of the tensor potential are sufficient to obtain equations of motion of the two sources till the first PN order. The monopole and dipole moments, respectively, of various components of $ \Lambda^{\mu\nu} $ (or $ \tilde{\Lambda}^{\mu\nu} $) are defined as follows
	\begin{eqnarray}
	P_L^{s} &=& \lim\limits_{\epsilon\,\rightarrow\,0}\, \int_{\mathcal{B}_L} d^3X_L \, \Lambda^{s\,s} \label{monoss}\\
	P^i_L &=& \lim\limits_{\epsilon\,\rightarrow\,0}\, \int_{\mathcal{B}_L} d^3X_L \, \Lambda^{s\,i} \label{monosi} \\
	Z^{i\,j}_L &=& \lim\limits_{\epsilon\,\rightarrow\,0}\, \int_{\mathcal{B}_L} d^3X_L \, \Lambda^{i\,j} \label{monoij}\\
	D^i_L &=& \lim\limits_{\epsilon\,\rightarrow\,0}\, \int_{\mathcal{B}_L} d^3X_L \, \Lambda^{s\,s} \, X^i_L \label{diss}\\
	J^{i\,j}_L &=& \lim\limits_{\epsilon\,\rightarrow\,0}\, \int_{\mathcal{B}_L} d^3X_L \, \Lambda^{s\,i} X^j_L \label{disi} \\
	Z^{i\,j\,k}_L &=& \lim\limits_{\epsilon\,\rightarrow\,0}\, \int_{\mathcal{B}_L} d^3X_L \, \Lambda^{i\,j} \, X^k_L \label{diij}
	\end{eqnarray}
	where $ \mathcal{B}_L $ implies the body zone/co-moving+scaled coordinate system of the $ L-th $ body zone.\\
	
	One defines a 4-momentum, by putting together the scalar in Eq. (\ref{monoss}) and the 3-vector in Eq. (\ref{monosi}), as \cite{Futamase1987,Itoh2000,Futamase2007,Giulini:2018}
	\begin{eqnarray}
	P^\mu_L\left(s\right) &=& \epsilon^2\,\int_{\mathcal{B}_L}\,d^3X_L\,\Lambda^{s\,\mu} \label{fourmomentum}
	\end{eqnarray}
	which along with the conservation law (\ref{EMconspartial}) or (\ref{effemcons}), leads to the following 4-force or the 4-momentum evolution formula \cite{Futamase1987,Itoh2000,Futamase2007}
	\begin{eqnarray}
	\frac{dP^\mu_L}{ds} &=& -\epsilon^{-4}\,\oint_{\partial\mathcal{B}_L}\,dS_k\,\Lambda^{k\,\mu} + \epsilon^{-4}\,v^k_L\,\oint_{\partial\mathcal{B}_L}\,dS_k\,\Lambda^{s\,\mu}, \label{momevol}
	\end{eqnarray}
	When one defines the bare quasi-local mass of the body zone $ L $ to be Eq. (\ref{monoss}), the 3-momentum to 3-velocity relation is given at the leading order by 
	\begin{eqnarray}
	P^i_L &=& P^s_L\,v^i_L + Q^i_L + \mathcal{O}\left(\epsilon^2\right) \\
	Q^i_L &=& \epsilon^{-4} \oint_{\partial\mathcal{B}_L} dS_k \left(\Lambda^{s\,k} - v_L^k\, \Lambda^{s\,s}\right) X_L^i \label{qi}
	\end{eqnarray}
	By choosing the origin of the coordinate system to be the center of mass of the body zone $ L $ (which is a choice that has already been made), $ Q^i_L $ can be shown to vanish at the leading (Newtonian) order, that is $ \mathcal{O}\left(1\right) $. Therefore, one obtains the 3-velocity evolution relation (or the 3-force on the body zone $ L $) as
	\begin{eqnarray}\label{vevol}
	P^s_L \, \frac{dv^i_L}{ds} &=& -\epsilon^{-4} \oint_{\partial\mathcal{B}_L} dS_k\, \Lambda^{k\,i} + \epsilon^{-4} \, v^k_L \oint_{\partial\mathcal{B}_L}dS_k \,\Lambda^{s\,i} \,+\, \epsilon^{-4}\,v^i_L \left(\oint_{\partial\mathcal{B}_L}dS_k\, \Lambda^{k\,s} \right.\nonumber\\
	&&\left. - v_L^k \oint_{\partial\mathcal{B}_L}dS_k \,\Lambda^{s\,s}\right) \,+\, \mathcal{O}\left(\epsilon^2\right),
	\end{eqnarray}
	All of the relations in this section (as mentioned along the way) holds both for GR and $ f\left(R\right) $ theories of gravity, and can be utilized to obtain the 2-body equations of motion of both theories, simply owing to the fact that the second Bianchi identity (or its generalization) holds for both theories of gravity. In GR for example, the equations of motion till 1 PN or $ \mathcal{O}\left(\epsilon^2\right) $ of Eq. (\ref{vevol}) for one of the bodies (say body zone 1) is given by
	\begin{eqnarray}\label{GReom}
	M_1\,\frac{dv_1^i}{ds} &=& -\frac{M_1\,M_2}{r_{1\,2}^{\,2}}\,n^i\,+\,\epsilon^2\,\frac{M_1\,M_2}{r_{1\,2}^2}\,\left[\left(-v_1^2-2\,v_2^2\,+\,\frac{3}{2}\,\left(\hat{n}\,\cdot\textbf{v}_2\right)^2+4\,\left(\textbf{v}_1\,\cdot\,\textbf{v}_2\right) \right.\right.\nonumber\\
	&&\left.\left.\,+\,\frac{5\,M_1}{r_{1\,2}} \,+\, \frac{4\,M_2}{r_{1\,2}}\right)\,n^i\,+\,\left\{4\,\left(\hat{n}\,\cdot\,\textbf{v}_1\right)\,-3\,\left(\hat{n}\,\cdot\,\textbf{v}_2\right)\right\}\,\left(v_1^i-v_2^i\right)\right]\,+\,\mathcal{O}\left(\epsilon^4\right)\nonumber\\
	\end{eqnarray}
	\section{Post Newtonian sequence of solutions and equations of motion for the quadratic deviation to the Einstein-Hilbert action}
	\subsection{The curious case of the homogeneous solution}
	One may be tempted to immediately go for the particular solution of the differential equation (\ref{scaldofdyn}), that is by solving for the following inhomogeneous problem using the retarded Green's function $ G_\gamma\left(x^\mu,\,x^{\mu'}\right) $
	\begin{eqnarray}
	\left(\Box\,+\,\sqrt{-g}\,\gamma^2\right)\,G_\gamma\left(x^\mu,x^{\mu'}\right) &=& \delta\left(x^\mu\,-\,x^{\mu'}\right) \label{kggreenfunction}
	\end{eqnarray}
	The support of the corresponding integral problem (that is the limit for $ x^{\mu'} $ in the Green's function integral) of the above lies in the past light-cone of the event $ x^\mu $. However, one must remember that the homogeneous solution of GR leads to the BH solutions. Hence, one must also consider the homogeneous solution of the Klein-Gordon equation, which must be solved simultaneously along with the homogeneous solution of GR.\\
	
	The homogeneous Klein-Gordon equation of the current theory under study, under the transformation (\ref{newtime}), has an $ \epsilon $ dependence given as follows
	\begin{eqnarray}
	\Box\mathfrak{R}_{hom} \,+\, \frac{\gamma^2}{\epsilon}\,\mathfrak{R}_{hom} &=& 0
	\end{eqnarray}
	While the radiative solutions of the above, under the choice of no-incoming scalar radiation from past null infinity, can be put to zero, the time independent part of the homogeneous solution cannot be ignored. The time independent differential equation is given by
	\begin{eqnarray}\label{homokg}
	\nabla\,\mathfrak{R}_{hom}\,-\,\frac{\gamma^2}{\epsilon}\,\,\mathfrak{R}_{hom}\,&=&\,0
	\end{eqnarray}
	where $ \nabla $ is the Laplace operator in Minkowski space-time. In spherical symmetry, for example, the LHS of Eq. (\ref{homokg}) is given by
	\begin{eqnarray}\label{kghomo}
	\frac{1}{r^2}\,\left[r^2\,\left(\mathfrak{R}_{hom}\right)_{,r}\right]_{,r}\,-\,\frac{\gamma^2}{\epsilon}\,\mathfrak{R}_{hom} &=& 0
	\end{eqnarray}
	which has a solution which is both regular at $ r\,=\,0 $ and $ r\,=\,\infty $, and is given by
	\begin{eqnarray}\label{riccihom}
	\mathfrak{R}_{hom} &=& \frac{C\,\,e^{-\frac{\gamma\,r}{\sqrt{\epsilon}}}}{r}
	\end{eqnarray}
	where $ C $ is a real constant of integration yet to be fixed. It is to be noted that the homogeneous solution, which may be part of an extended homogeneous solution of the combined scalar+tensor theory under study (like the Schwarzschild and Kerr solutions of GR), for vanishing post Newtonian or post Minkowskian parameter $ \epsilon $, leads to an infinite mass of the scalar field. Which also implies for low energies, the Ricci scalar or the Ricci scalar density considered here is practically un-excitable.\\
	
	However, it is interesting to look at the effect of $ \mathfrak{R} $ on the tensor wave equation (\ref{fRLL}), and how it modifies the Newtonian order solution, or more specifically, the Newtonian potential. The time-time component of the conformally transformed metric density deviation satisfies the flat-space wave equation with a source, following Eq. (\ref{LLfR})
	\begin{eqnarray}
	\Box\tilde{\mathfrak{h}}^{t\,t} &=& 16\,\pi\,\left(-g\right)\,T^{t\,t}\,\left(1\,-\,\epsilon\,f''\,\mathfrak{R}\right) \label{ttwave}
	\end{eqnarray}
	where the leading order, or only the terms in Eq. (\ref{effEMSfR}) that are proportional to the classical or matter energy-momentum-stress density tensor were used, along with Eq. (\ref{scaleddetmetric}). The above can be solved by using the retarded Green's function for the massless wave operator, with $ \tilde{\mathfrak{h}}^{s\,s} $ given by the body zone integral
	\begin{eqnarray}
	\tilde{\mathfrak{h}}^{s\,s} &=& 4\,\epsilon^4\,\sum_{L=1,\,2}\,\int_{\mathcal{B}_L}\,d^3X_L\,\frac{T^{s\,s}\,\left(1\,-\,\epsilon\,f''\,\mathfrak{R}\right)}{\left|Z^i_L\,-\,\epsilon^2\,X^i_L\right|}
	\end{eqnarray}
	Eq. (\ref{scaleddetmetric}) and (\ref{emscalss}) was used to obtain the above from Eq. (\ref{ttwave}).	The first integral, and its $ \epsilon $ expansion of the series is given in Eq. (\ref{fhssmul}), and is the usual Newtonian potential. However, a more curious thing occurs when one considers the extra term coming due to the quadratic modification to the gravitational action
	\begin{eqnarray}
	\tilde{\mathfrak{h}}^{s\,s}_{extra} &=& -4\,\epsilon^5\,f''\,\sum_{L=1,\,2}\,\int_{\mathcal{B}_L}\,d^3X_L\,\frac{T^{s\,s}\,\mathfrak{R}_{hom}}{\left|Z^i_L\,-\,\epsilon^2\,X^i_L\right|}
	\end{eqnarray}
	Considering now the particular form of the homogeneous Ricci density (\ref{riccihom}), one obtains the following integral using the scaled coordinate (\ref{scalcoord})
	\begin{eqnarray}
	\tilde{\mathfrak{h}}^{s\,s}_{extra} &=& -4\,\epsilon^3\,f''\,\sum_{L=1,\,2}\,C_L\,\int_{\mathcal{B}_L}\,d^3X_L\,\frac{T^{s\,s}}{\left|Z^i_L\left(s\right)\,-\,\epsilon^2\,X^i_L\right|\,\left|X^i_L\right|}\,e^{-\gamma\,\epsilon^{\frac{3}{2}}\,\left|X^i_L\right|}
	\end{eqnarray}
	which seems to preceed the order at which the gothic metric deviation first appears, that is at $ \mathcal{O}\left(\epsilon^3\right) $ compared to $ \mathcal{O}\left(\epsilon^4\right) $. However, expanding the extra term about $ \epsilon\,\rightarrow\,0 $ leads to the following
	\begin{eqnarray}
	\tilde{\mathfrak{h}}^{s\,s}_{extra} &=& -4\,\epsilon^3\,f''\,\sum_{L=1,\,2}\,\frac{C_L}{\left|Z^i_L\right|}\,\int_{\mathcal{B}_L}\,d^3X_L\,\frac{T^{s\,s}}{\left|X^i_L\right|}\\
	\tilde{\mathfrak{h}}^{s\,s}_{extra} &=& -4\,\epsilon^3\,f''\sum_{L=1,\,2}\,\frac{C_L\,N_{L}}{\left|X^i_L\right|}
	\end{eqnarray}
	which is a Newtonian potential like term with negative one multipole moment of the source, denoted by $ N_L $. Therefore, the modified leading order metric deviation $ \tilde{\mathfrak{h}}^{s\,s} $ can be written in the following form
	\begin{eqnarray}
	\tilde{\mathfrak{h}}^{s\,s} &=& 4\,\epsilon^4\,\sum_{L=1,\,2}\,\left[\frac{\tilde{P}^s_L}{\left|Z^i_L\right|}\,+\,\mathcal{O}\left(\epsilon^2\right)\right]\\
	\tilde{P}^s_L &=& P^s_L\,-\,\lim\limits_{\epsilon\,\rightarrow\,0}\,\frac{f''\,C_L}{\epsilon}\,\int_{\mathcal{B}_L}\,\,d^3X_L\,\frac{T^{s\,s}}{\left|X^i_L\right|} \label{massdefdiv}
	\end{eqnarray}
	where $ P^s_L $ was defined in Eq. (\ref{monoss}). It is important to note that while the integral in the second term of the square bracket of the above has a well defined limit for $ \epsilon\,\rightarrow\,0 $, the coefficient multiplying it does not. In fact, due to the $ \epsilon^{-1} $ nature of the coefficient, the modified mass-energy definition blows up at the $ \epsilon\,\rightarrow\,0 $ limit, leading to the following conclusion
	\begin{eqnarray}
	C_L &=& 0,
	\end{eqnarray}
	implying that the homogeneous Ricci scalar does not modify the tensor deviation density at all.
	\subsection{Deviation from the general relativistic equations of motion}
	The leading order Ricci scalar density deviation occurs at $ \mathcal{O}\left(\epsilon^4\right) $, and is given by the following
	\begin{eqnarray}\label{ricciinhom}
	{}_{(4)}\mathfrak{R}_{part} &\equiv& {}_{(4)}\mathfrak{R} \,=\, -8\,\pi\,\gamma^2\,\epsilon^4\,\frac{M_L\,\,e^{-\frac{\gamma\,\left|Z^i_L\right|}{\sqrt{\epsilon}}}}{\left|Z^i_L\right|}
	\end{eqnarray}
	whose form is the same as the now extinct homogeneous solution, but instead of the unknown constant, there is just the mass coming as a parameter, as in the definition of Eq. (\ref{monoss}). The only unknown in the above equation is the inverse length scale $ \gamma $, which was related to the coefficient of the quadratic deviation of the GR action in Eq. (\ref{gammadef}). Derivation of the above has been given in Appendix A.\\
	
	In GR, the non-vanishing and body zone independent terms in the surface integrals of Eq. (\ref{vevol}) lead to the 3-force at different PN orders, and has been well documented in PN literature, for example \cite{Itoh:2001np,Itoh:2003fz}. However, in this study only the leading order modification to the general relativistic equations of motion due to the quadratic deviation to the GR action will be quoted. Such a modification can be found by using the modified conservation law (\ref{effemcons}) along with the definitions of (\ref{momevol}) and (\ref{vevol}) with $ \Lambda^{\mu\nu} $ replaced by $ \tilde{\Lambda}^{\mu\nu} $. In Eq. (\ref{vevol}), various components of the modified net energy-momentum-stress density tensor $ \tilde{\Lambda}^{\mu\nu} $ will appear as integrand of the surface integrals. However, the components that have a non-vanishing and \emph{body zone boundary radius} independent contribution to the surface integrals of Eq. (\ref{vevol}), was found to be coming from the space-space component of $ t_{eff}^{\mu\nu} $, or
	\begin{eqnarray}
	&&t^{i\,j}_{eff} = \frac{\epsilon^4}{192\,\pi\,\gamma^4}\,\left[2\,\left(\eta^{i\,k}\,\eta^{j\,l}\,+\,\eta^{i\,l}\,\eta^{j\,k}\,-\,\eta^{i\,j}\,\eta^{k\,l}\right)\,\mathfrak{R}\,\partial_k\mathfrak{R}\,\partial_l\tilde{\mathfrak{h}}\right.\nonumber\\
	&&\left.\,+\,\left(\eta^{i\,k}\,\eta^{j\,l}\,-\,\frac{1}{2}\,\eta^{i\,j}\,\eta^{k\,l}\right)\,\left(4\,\partial_k\mathfrak{R}\,\partial_l\mathfrak{R}\,+\,\mathfrak{R}^2\,\partial_k\tilde{\mathfrak{h}}\,\partial_l\tilde{\mathfrak{h}}\right)\right]\label{lambdaeff}
	\end{eqnarray}
	where Eq. (\ref{scaleddetmetric}) was used to replace the metric determinant, or $ \frac{1}{\left(-g\right)^2} $ multiplying the quadratic $ f'' $ part of $ t^{\mu\nu}_{eff} $ in Eq. (\ref{teffrel}), which appear as $ \epsilon^4 $ above.\\
	
	Eq. (\ref{lambdaeff}) has three different $ \epsilon $ orders worth of terms inside the square bracket. When one utilizes the scalings of the leading order deviation of either $ \tilde{\mathfrak{h}} $ or $ \mathfrak{R} $, one notices that $ \tilde{\mathfrak{h}}\,\sim\,\epsilon^2 $ and $ \mathfrak{R}\,\sim\,\epsilon^4 $, following Eqs. (\ref{scaledminkowski}), (\ref{fhssmul})-(\ref{fhijmul}), and (\ref{ricciinhom}). From the number of times either $ \tilde{\mathfrak{h}} $ and $ \mathfrak{R} $ appear inside the square bracket in each product of fields, the first term scale as $ \epsilon^{10} $, the second as $ \epsilon^8 $, and the last one as $ \epsilon^{12} $. Since the overall factor of $ \epsilon^4 $ cancels away due to $ \epsilon^{-4} $ multiplying each surface integral of Eq. (\ref{vevol}), the three terms inside the square brackets of Eq. (\ref{lambdaeff}) are 5th, 4th, and 6th PN effects on the two body equations of motion. However, not all of them survive the surface integral. Only the last term in the square bracket survives and is a 6th PN effect with the modification to the 3-force on the $ L^{th} $ body given as follows
	\begin{eqnarray}
	M_L\,\frac{dv^i_L}{ds} &=& F^i_{N+PN,GR} \,-\,\epsilon^{12}\,\frac{32\,\pi\,M_L\,M_{L'}^3}{9\,r_{1\,2}^4}\,\,e^{-\frac{2\,\gamma\,r_{1\,2}}{\sqrt{\epsilon}}}\,n^i \label{fRmodforce}
	\end{eqnarray}
	where $ M_L $ is the quasi-local mass enclosed by the $ L^{th} $ body zone boundary, and $ M_{L'} $ is the quasi-local mass enclosed by the partner object's body zone boundary in the two-body system. $ r_{1\,2} $ is the distance between the origin of the two body zones, whereas $ n^i $ is a unit vector pointing from the origin of body zone $ L $ to the origin of body zone $ L' $. The negative sign on the modified/extra force imply that the direction of the force is antiparallel to $ n^i $, which is then a repulsive modification to the usual attractive Newtonian and PN forces indicated by the first term in the RHS of the above, that is $ F^i_{N+PN,GR} $. Compared to other works in literature on $ f\left(R\right) $ theories that calculate the conserved potential (or the force) from modifications to the time-time component of the metric tensor, the above has an inverse quartic distance falloff (along with the Yukawa exponential falloff) at 6 PN order, which is an extremely tiny modification to the general relativistic PN equations of motion at the very high energy scales of ultra-compact two-body interactions.
	\section{Discussions and conclusion}
	In the current study a particular redefinition of the standard gothic metric was utilized to drastically simplify the field equations of general $ f\left(R\right) $ theories of gravity. Once the identification was made that $ f'\left(R\right) $ acts like a massive scalar field, a fourth order system was reduced to two second order system of differential equations. The field equations were framed in the densitized approach following the methods of Landau and Lifshitz, and their redefinition of the field equations of GR. The use of conformal densities to define Landau-Lifshitz like field equations were used in scalar-tensor theories, for example in \cite{Mirshekari:2013vb,Kopeikin:2020rxp}. In this study, it is shown that the redefinition to a conformal density greatly suppresses the actual effect of an extra massive scalar degree of freedom on orbital dynamics.\\
	
	Following the definition of a tensor density, a scalar density $ \mathfrak{R} $, corresponding to the massive scalar degree of freedom was defined, which relates to the trace of the energy-momentum-stress density tensor in the same manner as the conformal metric tensor density relates to the energy-momentum-stress density tensor in the Landau-Lifshitz like tensor field equations of $ f\left(R\right) $ theories. The differential equation that is satisfied by $ \mathfrak{R} $ throws up a static solution that makes the mass-energy definition at the Newtonian/leading order diverge, leading to its removal from further usage. At the leading order, the particular/inhomogeneous $ \mathfrak{R} $ is seen to be static as well, and has the Yukawa potential form with a cutoff/inverse length scale that is related to the coefficient of the quadratic modification (with $ \left(\text{length}\right)^2 $ dimensions) to the Einstein-Hilbert action.\\
	
	However, just finding out the particular form of the leading order deviation of $ \mathfrak{R} $ is quite useless, unless one finds its corresponding effect on the conformal tensor density field equations, and then, on the PN equations of motion. The leading order $ \mathfrak{R} $ appears as an effective source tensor that drives the behavior of $ \tilde{\mathfrak{h}}^{\mu\nu} $, as well as the evolutionary dynamics of the quasi-local four-momentum (\ref{momevol}), and correspondingly, the velocity evolution equations (\ref{vevol}) through surface integrals of various components of the energy-momentum-stress density tensor. It was found in Eq. (\ref{fRmodforce}) that the modification to the GR PN equations of motion, due to quadratic modification at the action level, appear at the sixth PN order. On top of that, the force has an inverse fourth powered dependence on the distance along with a Yukawa falloff, which is a very tiny modification to GR equations of motion. The modified force is dependent on the masses enclosed by the body zone boundaries, and the modification to the equations of motion of one of the objects is independent of the mass enclosed in that object's body zone boundary - implying that the equivalence principle (as far as it holds in the GR PN scheme) is not affected by the quadratic modification to GR. The modified force is also conservative, and hence will not be adding to any extra dissipation (other than GR) of the orbital energy of the binary system, although it might lead to very tiny modifications to the phase evolution of the binary, as encoded in the gravitational waves. More importantly, the force is repulsive, compared to the PN forces that appear on perturbing GR about Minkowski space-times. So in the course of the binary evolution, at length scales comparable to $ \gamma^{-1} $, if the two objects have not merged into a single object already, they will feel a repulsive force that will compete with the attractive GR PN forces coming out of the \emph{superpotentials} of the PN expansion of GR, which are themselves progressively short ranged. Putting a constraint on the new length scale $ \gamma^{-1} $ is beyond the scope of this study, mostly because the author is not aware of sixth PN equations of motion appearing in the literature.
	
	\appendix\\
	The net solution to the  Klein-Gordon problem (\ref{scaldofdyn}) with the boundary condition choice of no incoming scalar radiation from past null Minkowskian infinity is given as follows
	\begin{eqnarray}\label{greenint}
	\mathfrak{R}\left(x^\mu\right) &=& -8\,\pi\,\gamma^2\,\epsilon^{-2}\,\int\,d^4y\, G_{\tilde{\gamma}}\left(x^\mu,\,y^{\mu}\right)\,T_m\left(y^{\mu}\right)\\
	\tilde{\gamma}^2 &\equiv& \frac{\gamma^2}{\epsilon}
	\end{eqnarray}
	where the factor of $ \epsilon^{-2} $ arises because the metric determinant $ \left(-g\right) $ multiplying the RHS of Eq. (\ref{scaldofdyn}), under the transformation (\ref{newtime}), scale as $ \epsilon^{-2} $, as seen after Eq. (\ref{scaledminkowski}). $ G_{\tilde{\gamma}}\left(x^\mu,\,y^\mu\right) $ is the retarded Green's function of the Klein-Gordon equation, as was given in \cite{Berry2011} with the $ \left(+\,,\,-\,,\,-\,,\,-\right) $ metric signature as
	\begin{eqnarray}\label{kgreen}
	G_{\tilde{\gamma}}\left(t,\,q;\,x^i,\,y^i\right) &=& \int_{-\infty,\,\tilde{\gamma}}^{-\tilde{\gamma},\,\infty} \frac{d\,\omega}{2\,\pi}\,e^{-i\,\omega\left(t-q\right)}\,\frac{e^{i\,\sqrt{\omega^2-\tilde{\gamma}^2}\,\left|x^i-y^i\right|}}{4\,\pi\,\left|x^i-y^i\right|} \nonumber\\
	&&+\,\int_{-\tilde{\gamma}}^{\tilde{\gamma}}\,\frac{d\omega}{2\,\pi}\,e^{-i\,\omega\,\left(t-q\right)}\,\frac{e^{-\,\sqrt{\tilde{\gamma}^2-\omega^2}\,\left|x^i-y^i\right|}}{4\,\pi\,\left|x^i-y^i\right|}\nonumber\\
	\end{eqnarray}
	The notation $ \int_{-\infty,\,\tilde{\gamma}}^{-\tilde{\gamma},\,\infty} $ involve two integrals, one from $ -\infty $ to $ -\tilde{\gamma} $, and the other from $ \tilde{\gamma} $ to $ \infty $. It is to be noted that the three integrals whose domains encompass all of $ \omega $ space must be evaluated simultaneously in order for the solution to converge. $ T_m\left(q,\,y^i\right) $, the trace of the classical matter energy-momentum tensor, which being a scalar, does not transform under any of the coordinate scalings.\\
	
	Due to the the coordinate scalings in Eq. (\ref{scalcoord}), the infinitesimal 4-volume element $ d^4y $ transform from the asymptotic observer's frame to either of the body zones $ \mathcal{B}_L $ in the following manner
	\begin{eqnarray}
	d^4y &\equiv& dt\,\wedge\,d^3y \\
	&\rightarrow& \epsilon^{-1}\,ds\,\wedge\,\epsilon^6\,\,d^3X_L \label{volelbod}
	\end{eqnarray}
	where $ \wedge $ denotes the wedge product between 1-form dt and 3-form $ d^3y $. In the current operational context $ \wedge $ is effectively scalar multiplication.\\
	
	Substituting Eq. (\ref{kgreen}) in Eq. (\ref{greenint}), and transforming into the body zone coordinates by substituting Eq. (\ref{volelbod}) as the infinitesimal covariant volume element in Eq. (\ref{greenint}), one obtains the following integral for the particular solution $ \mathfrak{R}_{part} $ for the inhomogeneous Klein-Gordon equation
	\begin{eqnarray}\label{rstep1}
	\mathfrak{R}_{part}\left(t,\,x^i\right) &=& -\epsilon^3\,\,8\,\pi\,\gamma^2\,\int_{-\infty}^{\infty}\,ds'\,\int_{\mathcal{B}_L}\,d^3X_L\,\,\int_{-\infty,\,\tilde{\gamma}}^{-\tilde{\gamma},\,\infty}\,\frac{d\omega}{2\,\pi}\,e^{-i\,\omega\,\left(t-s'/\epsilon\right)}\nonumber\\
	&& \times\quad\,\frac{e^{i\,\sqrt{\omega^2-\tilde{\gamma}^2}\,\left|Z^i_L-\epsilon^2\,X^i_L\right|}}{4\,\pi\,\left|Z^i_L-\epsilon^2\,X^i_L\right|}\,\left[\epsilon^4\,T_m\left(s'/\epsilon,\,X_L^i\right)\right]\nonumber\\
	&&-\,\epsilon^3\,\quad\,8\,\pi\,\gamma^2\,\int_{-\infty}^{\infty}\,ds'\,\int_{\mathcal{B}_L}\,d^3X_L\,\int_{-\tilde{\gamma}}^{\tilde{\gamma}}\,\frac{d\omega}{2\,\pi}\,e^{-i\,\omega\,\left(t-s'/\epsilon\right)}\nonumber\\
	&&\times\quad\,\frac{e^{-\sqrt{\tilde{\gamma}^2-\omega^2}\,\left|Z_L^i-\epsilon^2\,X^i_L\right|}}{4\,\pi\,\left|Z_L^i-\epsilon^2\,X^i_L\right|}\,\left[\epsilon^4\,T_m\left(s'/\epsilon,\,X^i_L\right)\right]\nonumber\\
	\end{eqnarray}
	
	In the body zones that surround each of the objects, the energy-momentum tensor of classical matter $ \left(T_{\mathcal{B}_L}\,=\,\lim\limits_{\epsilon\,\rightarrow\,0}\epsilon^4\,T_m\right) $ will be assumed to have a quasi-stationary/adiabatic initial condition for solving the relaxed system of equations, in the absence of any other time dependent driving force inside the compact object, and is given by
	\begin{eqnarray}\label{sauceass}
	T_{\mathcal{B}_L}\left(X^i\right) &=&  \sum_{n=-\infty}^{\infty}\,\mathfrak{T}_n\left(\left|X^i_L\right|,X^\theta_L\right)\,e^{i\,n\,\mathcal{X}}\\
	\mathcal{X}\left(s/\epsilon,\,X_L^\phi\right) &\equiv& \mathcal{X}\,=\,X^\phi_L\,-\,\Omega\,s/\epsilon\,+\,\phi_0 
	\end{eqnarray}
	where an axial+time symmetry was assumed for the initial condition, such that the time dependence of the source body in the body zone coordinate system repeats after every $ T\,=\,\frac{2\,\pi}{\Omega} $, with a constant phase parameter $ \phi_0 $. The weighing factors $ \mathfrak{T}_n $ are coefficients in the series expansion of the trace $ T_m \left(\text{or } T_{\mathcal{B_L}}\right) $ using stationary functions $ e^{i\,n\,\mathcal{X}} $, and are functions of the radial and azimuthal coordinates. Eq. (\ref{sauceass}) physically implies that the variations in the energy-momentum tensor sourcing $ \mathfrak{R} $ in the body zone of the first object is purely generated by the effect of the motion of the second object around it. The body zone coordinates $ \left(X^i_L\right) $ were defined in Eq. (\ref{scalcoord}), in which $ \left|X^i_L\right| $ is the distance from the center of mass of the body $ L $ to any point in the body zone coordinates, as viewed in the respective body zones. The choice in Eq. (\ref{sauceass}) has a simplifying effect on the subsequent calculations and is justified by the adiabatic and stationarity in the co-moving frame approximations, as found in the literature on PN expansions. \\
	
	Assuming quasi-periodicity for the source tensor, as in Eq. (\ref{sauceass}), one can substitute it into Eq. (\ref{rstep1}), to obtain
	\begin{eqnarray}
	\mathfrak{R}_{part}\left(t,\,x^i\right) &=& -\epsilon^3\,\,\frac{8\,\pi\,\gamma^2}{2\,\pi}\,\sum_{n=-\infty}^{\infty}\,\int_{\mathcal{B}_L}\,d^3X_L\,\left(\int_{-\infty,\,\tilde{\gamma}}^{-\tilde{\gamma},\,\infty}\,d\omega\,e^{-i\,\omega\,t}\right)\,e^{i\,n\,\left(X^\phi_L+\phi_0\right)}\nonumber\\
	&&\times\quad\,\frac{e^{i\,\sqrt{\omega^2-\tilde{\gamma}^2}\,\left|Z^i_L-\epsilon^2\,X^i_L\right|}}{4\,\pi\,\left|Z^i_L-\epsilon^2\,X^i_L\right|}\,\mathfrak{T}_n\left(\left|X^i_L\right|,X^\theta_L\right)\,\int_{-\infty}^{\infty}\,ds'\,e^{i\,\left(\omega-n\,\Omega\right)\,s'/\epsilon}\nonumber\\
	&&\,-\,\epsilon^3\quad\,\frac{8\,\pi\,\gamma^2}{2\,\pi}\,\sum_{n=-\infty}^{\infty}\,\int_{\mathcal{B}_L}\,d^3X_L\,\left(\int_{-\tilde{\gamma}}^{\tilde{\gamma}}\,d\omega\,e^{-i\,\omega\,t}\right)\,e^{i\,n\,\left(X^\phi_L+\phi_0\right)}\nonumber\\
	&&\times\quad\,\frac{e^{-\,\sqrt{\tilde{\gamma}^2-\omega^2}\,\left|Z^i_L-\epsilon^2\,X^i_L\right|}}{4\,\pi\,\left|Z^i_L-\epsilon^2\,X^i_L\right|}\,\mathfrak{T}_n\left(\left|X^i_L\right|,X^\theta_L\right)\,\int_{-\infty}^{\infty}\,ds'\,e^{i\,\left(\omega-n\,\Omega\right)\,s'/\epsilon}\nonumber\\
	\end{eqnarray}
	Transforming
	\begin{eqnarray}
	s' \rightarrow \epsilon\,s
	\end{eqnarray}
	and using the integral representation of the delta function,
	\begin{eqnarray}
	\int_{-\infty}^{\infty}\,e^{i\,k\,\left(x-x'\right)}\,dk &=& 2\,\pi\,\delta\left(x-x'\right),
	\end{eqnarray}
	one obtains
	%	\cleardoublepage
	\begin{eqnarray}\label{riccisum}
	\mathfrak{R}_{part}\left(t,\,x^i\right) &=& -\epsilon^4\,\,8\,\pi\,\gamma^2\,\sum_{n=-\infty}^{\infty}\,\int_{\mathcal{B}_L}\,d^3X_L\,\frac{e^{i\,\sqrt{n^2\,\Omega^2-\tilde{\gamma}^2}\,\left|Z^i_L-\epsilon^2\,X^i_L\right|}}{4\,\pi\,\left|Z^i_L-\epsilon^2\,X^i_L\right|}\nonumber\\
	&&\times\quad\,\mathfrak{T}_n\left(\left|X^i_L\right|,X^\theta_L\right)\,\times\,e^{-i\,n\,\left(\Omega\,t\,-\,X^\phi_L\,-\,\phi_0\right)}\\
	&&-\epsilon^4\,\,8\,\pi\,\gamma^2\,\sum_{n=-\infty}^{\infty}\,\int_{\mathcal{B}_L}\,d^3X_L\,\frac{e^{-\,\sqrt{\tilde{\gamma}^2-n^2\,\Omega^2}\,\left|Z^i_L-\epsilon^2\,X^i_L\right|}}{4\,\pi\,\left|Z^i_L-\epsilon^2\,X^i_L\right|}\nonumber\\
	&&\times\quad\,\mathfrak{T}_n\left(\left|X^i_L\right|,X^\theta_L\right)\,e^{-i\,n\,\left(\Omega\,t\,-\,X^\phi_L\,-\,\phi_0\right)}
	\end{eqnarray}
	
	The Green's function (\ref{kgreen}) can be written as an infinite sum of spherical harmonic functions, that are weighed by functions of the radial coordinates $ \left|Z^i_L\right| $ and $ \left|X^i_L\right| $, in the following way for $ \left|Z^i_L\right|\,>\,\left|X^i_L\right| $
	\begin{eqnarray}
	\frac{e^{i\,k\,\left|X^i\,-\,Z^i\right|}}{4\,\pi\,\left|X^i\,-\,Z^i\right|} &=& i\,k\,\sum_{\ell,\,m}j_\ell\left(k\,\left|X^i\right|\right)\,h_\ell^{(1)}\left(k\,\left|Z^i\right|\right)\,Y^*_{\ell\,m}\left(X^\theta,\,X^\phi\right)\,Y_{\ell\,m}\left(Z^\theta,\,Z^\phi\right) \nonumber\\\label{greenexpupper}\\
	\frac{e^{-\,k\,\left|X^i\,-\,Z^i\right|}}{4\,\pi\,\left|X^i\,-\,Z^i\right|} &=& \sum_{\ell,\,m}\,\frac{I_{\ell+\frac{1}{2}}\left(k\,\left|X^i\right|\right)\,K_{\ell+\frac{1}{2}}\,\left(k\,\left|Z^i\right|\right)}{\sqrt{\left|X^i\right|\,\left|Z^i\right|}}\,Y^*_{\ell\,m}\left(X^\theta,\,X^\phi\right)\,Y_{\ell\,m}\left(Z^\theta,\,Z^\phi\right)\nonumber\\\label{greenexplower}
	Z^i_L &\equiv& \frac{x^i\,-\,z^i_L\left(s\right)}{\epsilon^2}\,;\quad\,L\,=\,1,\,2.
	\end{eqnarray}
	The various functions appearing above are as follows
	\begin{itemize}
		\item $ j_\ell $: Spherical Bessel function of first kind.
		\item $ h_\ell^{(1)} $: Spherical Bessel function of third kind.
		\item $ I_{\ell\,+\,\frac{1}{2}} $: Modified Bessel function of the first kind.
		\item $ K_{\ell\,+\,\frac{1}{2}} $: Modified Bessel function of the second kind.
		\item $ Y_{\ell\,m},\,Y^*_{\ell\,m} $: Spherical harmonic functions and their complex conjugates.
	\end{itemize}
	
	After using Eqs. (\ref{greenexpupper}) and (\ref{greenexplower}) in the above, one obtains for $ \left|Z^i_L\right|\,>\,\left|X^i_L\right| $

	\begin{eqnarray}\label{riccipostexp}
		\mathfrak{R}_{part}\left(t,\,x^i\right) &=& -\epsilon^4\,\,8\,\pi\,i\,\gamma^2\,\sum_{n,\,\ell,\,m}^{\left|n\right|\,>\,\floor*{\frac{\tilde{\gamma}}{\Omega}}}\,\sqrt{n^2\,\Omega^2-\tilde{\gamma}^2}\,\int_{\mathcal{B}_L}\,d^3X_L\,e^{i\,n\,\left(X_L^\phi+\phi_0-\Omega\,t\right)}\nonumber\\
		&&\times\quad\,j_\ell\left(\epsilon^2\,\sqrt{n^2\,\Omega^2-\tilde{\gamma}^2}\,\left|X_L^i\right|\right)\,h_\ell^{(1)}\left(\sqrt{n^2\,\Omega^2-\tilde{\gamma}^2}\,\left|Z^i\right|\right)\nonumber\\
		&&\times\,\quad\,Y^*_{\ell\,m}\left(X_L^\theta,\,X_L^\phi\right)\,Y_{\ell\,m}\left(Z^\theta,\,Z^\phi\right)\,\mathfrak{T}_n\left(\left|X_L^i\right|,\,X_L^\theta\right)\nonumber\\
		&&-\,\epsilon^3\,\quad\,\,8\,\pi\,\gamma^2\,\sum_{n,\,\ell,\,m}^{\left|n\right|\,<\,\floor*{\frac{\tilde{\gamma}}{\Omega}}}\,\int_{\mathcal{B}_L}\,d^3X_L\,e^{i\,n\,\left(X_L^\phi+\phi_0-\Omega\,t\right)}\nonumber\\
		&&\times\quad\,\frac{I_{\ell+\frac{1}{2}}\left(\epsilon^2\,\sqrt{\tilde{\gamma}^2-n^2\,\Omega^2}\,\left|X_L^i\right|\right)}{\sqrt{\left|X_L^i\right|}}\,\frac{K_{\ell+\frac{1}{2}}\left(\sqrt{\tilde{\gamma}^2-n^2\,\Omega^2}\,\left|Z^i\right|\right)}{\sqrt{\left|Z^i\right|}}\nonumber\\
		&&\times\quad\,Y^*_{\ell\,m}\left(X_L^\theta,\,X_L^\phi\right)\,Y_{\ell\,m}\left(Z^\theta,\,Z^\phi\right)\,\mathfrak{T}_n\left(\left|X_L^i\right|,\,X_L^\theta\right)\nonumber\\
		\end{eqnarray}

	one can substitute the following properties of spherical harmonic functions in Eq. (\ref{riccipostexp})
	\begin{eqnarray}
	&&Y^*_{\ell\,m} = \left(-1\right)^m\,Y_{\ell\,-m}\\
	&&Y_{\ell\,-m} = N_{\ell\,-m}\,P_{\ell\,-m}\,\left(\cos\,X^\theta_L\right)\,e^{-i\,m\,X^\phi}\\
	&&\int_{0}^{2\,\pi}\,dX^\phi_L\,Y^*_{\ell\,m}\,e^{i\,n\,X^\phi} = 2\,\pi\,\left(-1\right)^m\,N_{\ell\,-m}\,P_{\ell\,-m}\left(X^\theta_L\right)\,\delta_{m\,n}\nonumber\\
	\label{intprop}\\
	&&N_{\ell\,m} = \left(-1\right)^m\,\sqrt{\frac{\left(2\,\ell\,+\,1\right)}{4\,\pi}\,\frac{\left(\ell\,-\,m\right)!}{\left(\ell\,+\,m\right)!}}
	\end{eqnarray} 
	where $ P_{\ell\,m}\left(X^\theta_L\right) $ are the associated Legendre polynomials and $ \delta_{m\,n} $ is the Kronecker delta distribution. Since $ m $ takes values between $ -\ell $ to $ \ell $, application of the Kronecker delta leads to the summation on $ \ell $ in the first term of Eq. (\ref{riccipostexp}) going from the lower integral part of $ \frac{\tilde{\gamma}}{\Omega} $ or $ \floor*{\frac{\tilde{\gamma}}{\Omega}} $ to $ \infty $, whereas the second terms' summation range takes $ \ell $ from zero to $ \floor*{\frac{\tilde{\gamma}}{\Omega}} $. The contributions to the particular solution from above the cutoff energy scale $ \tilde{\gamma} $ is highly suppressed (especially at lower PN orders) because of the properties of $ j_\ell $ and $ h^{(1)}_\ell $ for asymptotically large values of $ \ell $, and terms in $ \mathfrak{R} $ series containing the same can be approximated to zero at the current orders of calculation. Hence one can safely set the first summation in RHS of Eq. (\ref{riccipostexp}) to zero and consider only the second summation. Hence it is to be noted that the factors that depend on $ \epsilon $ in the above are only in the arguments of the Bessel function $ I_{\ell\,+\,\frac{1}{2}} $.\\
	
	Hence the only relevant factors that explicitly contain $ \epsilon $ in both the sums in Eq. (\ref{riccipostexp}), can be expanded about $ \epsilon\,\rightarrow\,0 $ as follows
	\begin{eqnarray}\label{bessprodexp}
	I_{\ell+\frac{1}{2}}\,\left(\epsilon^2\,\Lambda\,\left|X_L\right|\right) &\approx& \epsilon^{2\,\ell\,+\,1}\,\frac{\Lambda^{\frac{2\,\ell\,+\,1}{2}}}{2^{\ell\,+\,\frac{1}{2}}\,\Gamma\left(\ell\,+\,\frac{3}{2}\right)}\,\left|X_L\right|^{\ell\,+\,\frac{1}{2}}\nonumber\\
	\label{flexp}
	\end{eqnarray}
	The approximation of (\ref{bessprodexp}), when substituted in for the relevant factors of the RHS of Eq. (\ref{riccipostexp}), leads to the following polynomial series of $ \epsilon $ about $ \epsilon\,\rightarrow\,0 $
	\begin{eqnarray}\label{ricciexp}
	&&\mathfrak{R}_{part}\left(t,\,x^i\right) \approx -8\,\pi\,\gamma^2\,\sum_{L,\,\ell,\,m}^{\ell\,=\,\infty}\,\epsilon^{2\,\ell\,+\,4}\,\frac{\left(\tilde{\gamma}^2\,-\,m^2\,\Omega^2\right)^{\frac{2\,\ell\,+\,1}{4}}}{2^{\ell\,+\,\frac{1}{2}}\,\tilde{\gamma}\left(\ell\,+\,\frac{3}{2}\right)}\nonumber\\
	&&\times\,e^{i\,m\,\left(\phi_0\,-\,\Omega\,t\right)}\,\,\frac{K_{\ell\,+\,\frac{1}{2}}\left(\sqrt{\tilde{\gamma}^2\,-\,m^2\,\Omega^2}\,\left|Z^i_L\right|\right)}{\sqrt{\left|Z^i_L\right|}}\,Y_{\ell\,m}\left(Z^\theta_L,\,Z^\phi_L\right)\nonumber\\
	&&\times\,\mathfrak{M}_{L\,\ell\,m}\\
	&&\mathfrak{M}_{L\,\ell\,m} = 2\,\pi\,\left(-1\right)^m\,N_{\ell\,-m}\,\int_{\mathcal{B}_L}\left|X^i_L\right|^{2+\ell}\,\sin \left(X^\theta_L\right)\nonumber\\
	&&\times\quad P_{\ell\,-m}\left(\cos X^\theta_L\right)\,\mathfrak{T}_m\left(\left|X^i_L\right|,\,X^\theta_L\right)d\left|X^i_L\right|\,dX^\theta_L\nonumber\\
	\end{eqnarray}
	From Eq. (\ref{ricciexp}), one can immediately notice that the leading order deviation for the particular solution of the Ricci scalar for the $ L^{th} $ body zone comes at $ \mathcal{O}\left(\epsilon^4\right) $ for $ \ell\,=\,0 $ with $ \mathfrak{M}_{L\,0\,0}\,\equiv\,M_L $ being the mass monopole, and was found to be
	\begin{eqnarray}\label{ricciinhomogeneous}
	{}_{(4)}\mathfrak{R}_{part} &\equiv& {}_{(4)}\mathfrak{R} \,=\, -8\,\pi\,\gamma^2\,\epsilon^4\,\frac{M_L\,\,e^{-\frac{\gamma\,\left|Z^i_L\right|}{\sqrt{\epsilon}}}}{\left|Z^i_L\right|}
	\end{eqnarray}
	\bibliographystyle{ws-procs961x669}
	%\bibliography{references}

\end{document}